\documentclass[aps,prl,twocolumn,superscriptaddress]{revtex4-2}
\usepackage{amsmath,amssymb,amsfonts,float,graphics,epsfig,epstopdf,color,verbatim,tabularx,bm,multirow,appendix,wasysym}
\usepackage{amsthm}
\usepackage[utf8]{inputenc}
\usepackage[T1]{fontenc}
\usepackage{xcolor}
\usepackage{dsfont}
\usepackage{textcomp}
\usepackage{yfonts}
\usepackage{bm}
\usepackage{subfigure}
\usepackage{mathrsfs}
\usepackage{graphicx}
\usepackage{verbatim}
\usepackage{hyperref}
\usepackage{multirow}
\usepackage{braket}
\usepackage[normalem]{ulem}
\usepackage{tikz}
	\usetikzlibrary{calc}
	\usetikzlibrary{shapes.multipart}

\renewcommand{\vec}[1]{{\mathbf #1}}

\newcommand{\comments}[1]{}

\newcommand{\stkout}[1]
{\ifmmode\text{\sout{\ensuremath{#1}}}\else\sout{#1}\fi}

\begin{document}
\title{Finite-time Scaling of the surface special transition in a 3D classical Heisenberg model} 

\author{Dong-Xu Liu}
\affiliation{Department of Physics, School of Science and Research Center for Industries of the Future, Westlake University, Hangzhou 310030,  China}
\affiliation{Institute of Natural Sciences, Westlake Institute for Advanced Study, Hangzhou 310024, China}

\author{Zhe Wang}
\email{wangzhe90@westlake.edu.cn}
\affiliation{Department of Physics, School of Science and Research Center for Industries of the Future, Westlake University, Hangzhou 310030,  China}
\affiliation{Institute of Natural Sciences, Westlake Institute for Advanced Study, Hangzhou 310024, China}

\author{Shuai Yin}
\email{yinsh6@mail.sysu.edu.cn}
\affiliation{School
of Physics, Sun Yat-sen University, Guangzhou 510275, China}
\affiliation{Guangdong Provincial Key Laboratory of Magnetoelectric Physics and Devices, Sun Yat-sen University, Guangzhou 510275, China}

\author{Zheng Yan}
\email{zhengyan@westlake.edu.cn}
\affiliation{Department of Physics, School of Science and Research Center for Industries of the Future, Westlake University, Hangzhou 310030,  China}
\affiliation{Institute of Natural Sciences, Westlake Institute for Advanced Study, Hangzhou 310024, China}

\begin{abstract} We investigate nonequilibrium driven dynamics across the special surface phase transition in the three-dimensional classical Heisenberg model with open boundaries, where tuning the surface coupling gives access to an extraordinary-log boundary critical state characterized by logarithmic, rather than power-law, decay of correlations. Using Monte Carlo simulations, we realize four driving protocols: temperature heating and cooling across the special transition, and surface-coupling ramps from the ordinary and extraordinary-log critical states into the special point. For temperature-driven protocols, the surface order parameter obeys a generalization of the finite-time scaling (FTS) and the Kibble–Zurek mechanism. The central finding emerges when the system is driven from the extraordinary-log critical state: the large-rate scaling relation acquires a logarithmic factor and takes the novel form $M^{2}_{s}\propto R^{(1+\eta_{s})/r_{s}}[\log(LR^{1/\eta_{s}})]^{-q}$, where $R$ is the driving rate, $L$ the system size, $\eta_{s}$ the surface anomalous dimension, $r_{s}$ the scaling dimension of $R$, and $q$ the exponent governing the logarithmic boundary criticality. We derive this scaling form from the generalized FTS framework by incorporating the logarithmic initial-state memory, and we achieve excellent data collapse over a wide range of system sizes and driving rates. Our results establish that extraordinary-log boundary-states alter the nonequilibrium critical scaling, extending boundary FTS beyond conventional power-law initial conditions.

\end{abstract}
\date{\today}
\maketitle

\textit{\color{blue} Introduction.---} In the study of critical phenomena, the response of a system to an external drive (e.g., a temperature quench) is central to understanding non-equilibrium dynamics~ \cite{Kibble_1976_Topology,Zurek_1985_Cosmological,Dziarmaga_2005_Dynamics,Polkovnikov_2005_Universal,Zurek_2005_Dynamics}. The Kibble–Zurek mechanism (KZM) provides a universal framework for predicting the system's non-equilibrium critical dynamics and has become a widely applied tool in both classical and quantum phase transitions~\cite{Yates1998Vortex,Uhlmann_2007_Vortex,Uhlmann_2010_Symmetry,Uhlmann_2010_System,Campo2014Universality,Bumsuk_2019_Kibble,Li_2023_Probing}. The essence of KZM focuses on the formation of the topological defects after the quench~\cite{Maegochi_2022_Kibble,Du_2023_Kibble,weinberg_2025_Defects}, and its central insight is the competition between the system's internal timescale for equilibration and the external timescale imposed by the quench. In addition, the finite‑time scaling (FTS) theory provides a more general framework that extends the KZM to the scaling behavior of arbitrary physical quantities and different kinds of driving protocols~\cite{Zhong2005Dynamic,Zhong_2006_Probing,Gong_2010_Finite,Huang_2010_Finite,Kolodrubetz_2012_Nonequilibrium,Yin_2014_Nonequilibrium,Huang_2016_Kibble,Yin_2016_Scaling}. For linearly varying the distance to the critical point, FTS demonstrates that the external driving
imposes a characteristic time scale $\xi_{\text{d}} \sim R^{-z/r}$ induced by the quench rate $R$, where $z$ is the dynamic critical exponent, $\nu$ the correlation‑length critical exponent, and $r = z + 1/\nu$ the scaling dimension of $R$. Recent works have further extended FTS to phase transitions beyond the Landau paradigm~\cite{Huang_2020_Kibble,Shu_2025_Equilibration,Zeng_2025_Finite,Zeng_2025_Nonequilibrium}, critical dynamics with multiple length scales~\cite{Shu_2024_Relaxation,Shu_2026_Finite}, dynamical phase transitions~\cite{Li_2019_Driving}, and its integration with relaxation critical dynamics~\cite{Huang_2016_Kibble}, thereby extending the KZM beyond adiabatic initial conditions~\cite{Zhong_2006_Probing,Huang_2016_Kibble,Yin_2016_Scaling}.

Investigations of boundary critical phenomena provide essential insight into how phase transitions manifest at interfaces, revealing a classification scheme that is significantly richer than that of their bulk counterparts~\cite{Chen_2012_Symmetry,Zhang_2017_Unconventional,Lukas_2019_Nonordinary,Zhu_2021_Surface,Wang_2022_Bulk,Wang_2023_Extraordinary}. Unlike bulk criticality, the  behavior is governed by the relative strength of surface interactions compared to bulk couplings, which leads to a tripartite division into three principal surface universality classes: ordinary transition, special transition and extra-ordinary transition~\cite{Binder_1974_Surface,Lubensky_1975_CriticalI,Lubensky_1975_CriticalII,Bray_1977_Critical,Diehl_1980_Scaling,Binder_1990_Critical,Gliozzi_2015_Boundary,Ding_2018_Engineering,Metlitski_2022_Boundary}. In the renormalization group picture, the special fixed point is unstable, controlling the crossover between the stable ordinary and extraordinary fixed points. In boundary conformal field theory, it corresponds to a non-trivial conformal boundary condition (often Robin-like) that merges the Dirichlet-type ordinary boundary with the symmetry-breaking extraordinary one~\cite{Diehl_2011_The,Jensen_2016_Constraint,Fursaev_2016_Anomalies,Herzog_2017_Boundary}. For systems with continuous symmetry, recent advances have revealed that the extraordinary transition itself can belong to the novel extraordinary-log universality class, where boundary correlation function decay logarithmically rather than as power laws~\cite{Hu_2021_Extraordinary,Toldin_2022_Boundary,Toldin_2025_Universal}, a feature that underscores the unique and subtle physics governing order at the boundary.

Against this background, applying non-equilibrium dynamics to the study of boundary criticality offers a promising approach to gaining deeper insight into boundary critical properties. For the relaxation dynamics, the dynamics of boundary criticality has been explored in the past decades~\cite{Dietrich_1983_The,Macoto_1985_Monte,Diehl_1994_Universality,Ritschel_1995_Universal,Michel_2004_Nonequilibrium,Michel_2005_Nonequilibrium,Lin_2008_Short}. In addition, Ref.~\cite{Shu_2025_Universal} pioneers the study of the driven critical dynamics near the boundary by generalizing the FTS theory to the boundary criticality and developing the boundary FTS (BFTS) theory to characterize the driven critical dynamics across the ordinary, special, and extraordinary transitions in the $3$D Ising~\cite{Shu_2025_Universal}. Although the edge critical effect of Ising criticality, as a textbook example, is sufficiently typical and representative, more complex and novel surface critical situations of driving dynamics remain completely unknown, such as the effects of the newly uncovered extraordinary-log transition~\cite{Toldin_2021_Boundary}.


When the bulk is at the 3D O(3) critical point, tuning the surface coupling can give rise to a particularly distinctive surface critical state known as the extraordinary-log phase, which is characterized by a logarithmic decay of correlations. With the extraordinary-log phase as the initial states, the logarithmic decay of the correlation function have the correlation length $\xi/L \sim [\log(L)]^{1/2}$~\cite{Toldin_2021_Boundary,Toldin_2022_Boundary,Toldin_2025_Universal}. It is quite instructive to explore the critical dynamics associated with the extraordinary-log phase.


This work investigates the driven critical dynamics of the classical 3D Heisenberg model with open boundaries along one direction. We focus on the critical driven-dynamic scaling behaviors near the boundary special transition point. As the intersection of the ordered phase at low temperature, disordered phase at high temperature, ordinary critical phase with small coupling and extraordinary-log phase with large coupling, this special point can provide abundant information on driven criticality. In detail, this unique configuration allows us to investigate four distinct driving protocols: heating and cooling across this critical point, as well as tuning the surface coupling to drive the system from either the ordinary or the extraordinary-log critical state toward the critical point. 
In the present case, the heating and cooling dynamics in such surface phase transitions are consistent with the conventional BFTS framework~\cite{Shu_2025_Universal}. 
More intriguingly, through both theoretical analysis and numerical verification, we reveal that an extraordinary-log critical initial state induces a novel scaling relation for the order parameter:
$ M^{2}_{s} \propto R^{(1+\eta_{s})/r_s}[\mathrm{log}(LR^{1/r_s})]^{-q} $.



\textit{\color{blue} Models and method.---} We simulate the 3D Heisenberg model on an $L_{\parallel} \times L_{\parallel} \times L$ lattice with $L_{\parallel}=L$, subject to periodic boundary conditions along the two $L_{\parallel}$ directions and open boundary conditions along the $L$ direction. The Hamiltonian of the system is given as follows:

\begin{equation}
\begin{split}
H= & -J \sum_{\langle ij\rangle} \vec{S}_{i} \cdot \vec{S}_{j} - J_{s,\uparrow}\sum_{\langle ij \rangle_{\uparrow}}\vec{S}_{i}\cdot\vec{S}_{i}  -J_{s,\downarrow}\sum_{\langle ij \rangle_{\downarrow}}\vec{S}_{i}\cdot\vec{S}_{i}
 \\& + \sum_{i} [\vec{S}^{2}_{i} + \lambda (\vec{S}^{2}_{i}-1)^{2}] 
\end{split}
\label{eq:sys_ham}
\end{equation}
where $\vec{S}_{i}$ is a 3D classical spin vector at lattice site $i$. The first term represents the total ferromagnetic interaction between all nearest-neighbor sites in the bulk. The second and third terms correspond to the ferromagnetic interactions between all nearest-neighbor sites on the upper and lower open boundary surfaces, respectively. The final term with $\lambda=5.2$ is introduced to suppress finite-size effects~\cite{Pelissetto_2002_Critical,Campostrini_2002_Critical,Hasenbusch_2020_Monte,Toldin_2021_Boundary}.


We set the temperature to unity $T=1$, corresponding to a bulk critical point at $J_c = 0.687985221$~\cite{Hasenbusch_2020_Monte,Toldin_2021_Boundary}. The symmetric surface couplings $J_{s,\uparrow}=J_{s,\downarrow}=J_{s}$ are employed to induce phase transitions on the boundary. By enhancing the surface coupling, the boundary sequentially enters an ordinary critical state characterized by algebraic decay of correlations, followed by an extraordinary-log critical state exhibiting logarithmic decay. The phase transition between these two regimes occurs at $J_{sc} = 1.16821$~\cite{Toldin_2021_Boundary} and is known as the special boundary phase transition.


The non-equilibrium evolution is simulated by the classical Monte Carlo (MC) method to study the dynamical scaling here. The process is implemented as multiple sequences of adiabatic pulses~\cite{Liu_2014_Dynamic,weinberg_2025_Defects}. Each sequence starts from an equilibrium state prepared in advance using Wolff cluster updates~\cite{Wolff_1989_Collective}. Within a sequence, each adiabatic pulse is realized by gradually varying a system parameter over a single Monte Carlo sweep, where one sweep involves $N=L^{3}$ flip attempts on randomly selected spins with the Metropolis acceptance criterion (i.e., single spin update). It has been established that the Metropolis dynamics falls within the Model A universality class, and is straightforward to implement in
experiments~\cite{Hohenberg1977Theory}. All sequences terminate at the same special critical point \cite{DeGrandi_2011_Universal,weinberg_2025_Defects}. As a sequence proceeds, the system parameter $P(t)$ evolves with the sequential step t:
$ P(t) = P_{0} + Rt$.
Here, $R$ is the driving rate, and $P$ represents a tunable parameter such as the bulk (surface) coupling  $J$ ($J_{s}$).  At the end of each sequences, we study the mean squared order parameter,
$M^{2}_{s} = \frac{1}{L_{||}^{4}}\langle (\sum_{i\in s} \vec{S}_{i})^2 \rangle$.



\begin{figure}[htp]
\centering
\includegraphics[width=1.0\columnwidth]{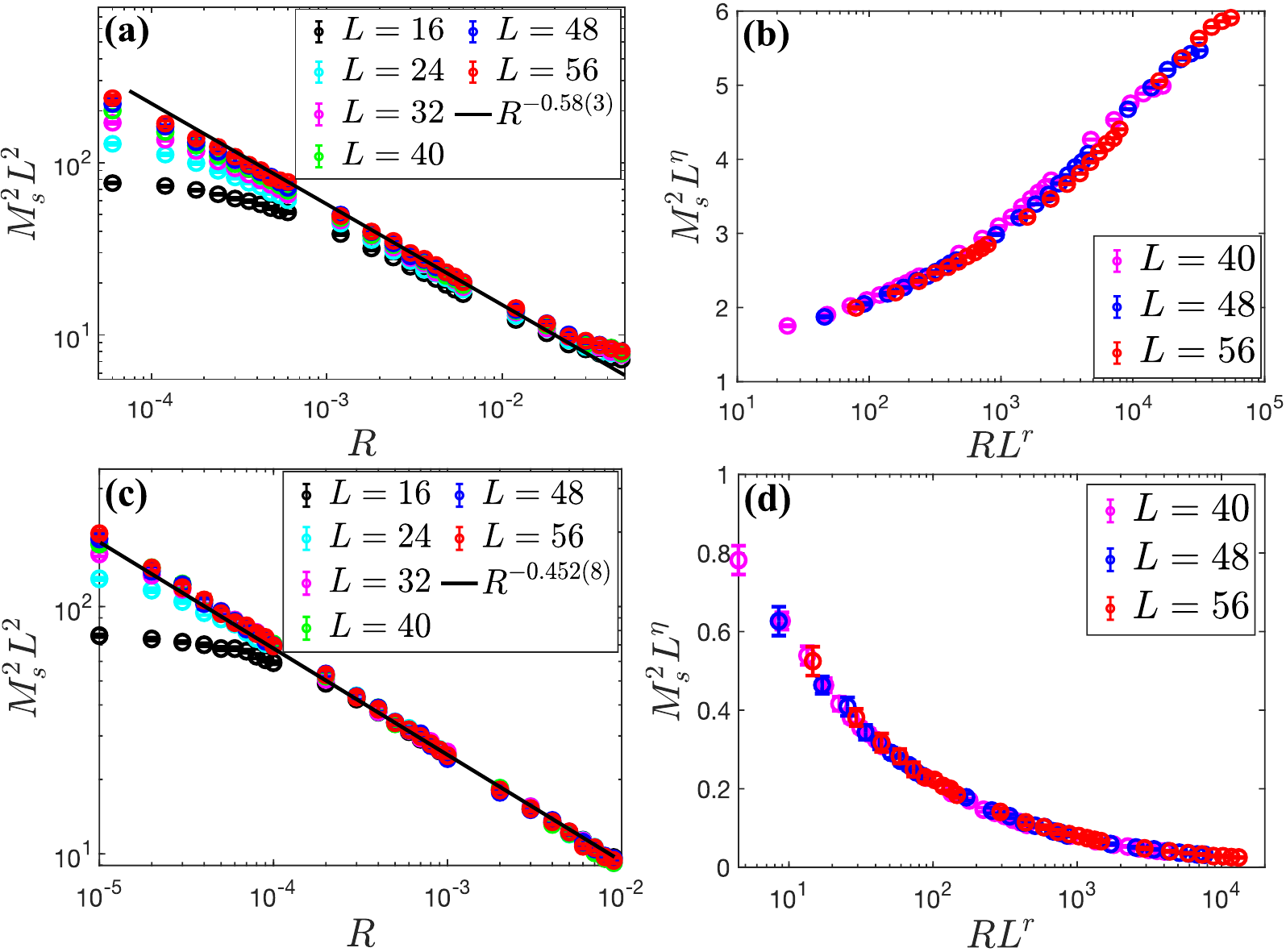}
\caption{Finite-time scaling analysis of the surface order parameter $M^{2}_{s}$ during non-equilibrium heating and cooling processes. (a) Heating process: the system is driven from an initial temperature $T_{0}=0.5$ to the special transition at $T_{t}=1.0$. The power law scaling behavior is represented by the black solid line. (b) After rescaling the $x$-axis and $y$-axis as shown in (b), a good data collapse is achieved for the non-equilibrium heating process when lower driving rate is employed. (c) Cooling process: the system is quenched from  $T_{0}=2.0$ to $T_{t}=1.0$. Its power law scaling behavior is indicated by the black solid curve. Data collapse analysis was performed on the non-equilibrium heating and cooling processes, both employing a low driving rate from their respective initial states. (d) After rescaling the $x$-axis and $y$-axis as shown in (d), a good data collapse in the non-equilibrium cooling process was achieved by employing low driving rate.}
\label{fig:cool_hat}
\end{figure}

\textit{\color{blue} Critical dynamics for changing the temperature.---}  
We investigate the temperature-driven heating and cooling dynamics across the special transition. The surface coupling is fixed at $J_{sc}=1.16821$. For the cooling process, the initial temperature is set to $T_{0}=2.0$, and the system is cooled to the special transition at temperature $T_{t}=1.0$. For the heating process, the initial temperature is $T_{0}=0.5$, and the system is heated to $T_{t}=1.0$. Fig.~\ref{fig:cool_hat} presents the driven dynamics of the surface order parameter $M^{2}_{s}$ during non-equilibrium heating (a) and cooling (c) processes. 

For the heating dynamics, Fig.\ref{fig:cool_hat}(a) shows how the surface order parameter $M^{2}_{s}$ evolves with the large driven rate $R$. The black solid line reveals a power-law scaling of $M^{2}_{s}$ with respect to the driving rate $R$. A power-law fit for system size $L=56$ yields the scaling relation $M^{2}_{s} \propto R^{0.186(5)}$. For the cooling dynamics, the power-law decay of the order parameter with the driving rate $R$ is shown in Fig.\ref{fig:cool_hat}(c). A fit for system size $L=56$, represented by the black solid line, yields the scaling relation $M^{2}_{s} \propto R^{-0.452(8)}$.

According to FTS theory, at $g=\vert T-T_{c}\vert=0$, $M^{2}$ satisfies the following scaling form~\cite{Zhong2005Dynamic,Huang2014Kibble,Feng2016Theory,Shu_2025_Universal}
\begin{equation}
\begin{split}
    M^{2}_{s} = L^{-(1+\eta_s)} \mathcal{M}\bigl(RL^{r}\bigr)
\end{split}
\label{eq:fit_eq}
\end{equation}
where $r=z+\frac{1}{\nu}$ denotes the critical dimensionality of  $R$ and $\eta_s=-0.473(2)$ denotes the anomalous dimension of the surface special phase transition~\cite{Toldin_2021_Boundary}  and $\mathcal{M}$ is the scaling function. For both heating and cooling processes, the correlation length is dominated by the bulk criticality. By incorporating the bulk critical exponents for the correlation length, $\nu \approx 0.707(3)$ \cite{Guillou_1989_Accurate,Cui_2008_Heisenberg}, and the dynamic critical exponent, $z \approx 2.033(3)$ \cite{Astillero_2019_Computation,Astillero_2025_Equilibrium}, we obtain $r \approx 3.447$. In the large-rate limit, the ordered initial state (heating) and the disordered initial state (cooling) yield distinct power-law scaling relations~\cite{Zhong2005Dynamic,Huang2014Kibble,Feng2016Theory,Shu_2025_Universal}: $M^{2}_{s} \propto R^{(1+\eta_s)/r} \approx R^{0.153}$ and $M^{2}_{s} \propto R^{(\eta_s-1)/r} \approx R^{-0.427}$, respectively. These theoretical predictions are in good agreement with our numerical results. Furthermore, we demonstrate that $M^{2}$ in both heating (Fig.\ref{fig:cool_hat}(b)) and cooling (Fig.\ref{fig:cool_hat}(d)) driven processes can be well rescaled using  Eq.~(\ref{eq:fit_eq}) (to minimize finite-size effects, data from the larger system sizes ($L = 40, 48, 56$) are employed in this analysis).  Consistent with Ref. \cite{Shu_2025_Universal}, we demonstrate that heating and cooling dynamics in such surface phase transitions can be described by the BFTS form.

\textit{\color{blue} Critical dynamics for changing the surface coupling.---}We investigate non-equilibrium dynamics by tuning the surface coupling $J_s$ from the ordinary critical state ($J_s=0.56821$) to the special transition point ($J_s=1.16821$), as well as from the extraordinary-log critical state ($J_s=1.66821$) to the same special point. Near this critical point, the surface correlation length scales as $\xi_s \propto |J_s - J_{sc}|^{-\nu_s}$ with $\nu_s \approx 2.78$~\cite{Toldin_2021_Boundary}. In this case, scaling analysis demonstrates that the driving rate $R$ has the dimension $r_s = z + 1/\nu_s \approx 2.3927$~\cite{Shu_2025_Universal}.

In equilibrium, the relationship between the correlation length and system size in the ordinary critical state and the extraordinary-log critical state is given by $\xi/L \propto \text{const}$ and $\xi/L \propto [\log(L)]^{1/2}$ \cite{Toldin_2021_Boundary,Toldin_2022_Boundary,Toldin_2025_Universal}, respectively. Although distinct in form, both imply a divergent relaxation time in the thermodynamic limit. When driven dynamics originates from a regime where the equilibrium relaxation time is already divergent and subsequently crosses the critical point, defects are profusely generated throughout this entire initial regime, rather than being confined to the vicinity of the critical point. How critical states shape the universal scaling behavior of driven dynamics has recently attracted growing attention~\cite{Zeng_2025_Finite,Zeng_2025_Nonequilibrium}, particularly in the context of boundary criticality~\cite{Shu_2025_Universal}. In this work, we reveal how such boundary critical states modulate boundary driven dynamics.

\begin{figure}[htp]
\centering
\includegraphics[width=1.0\columnwidth]{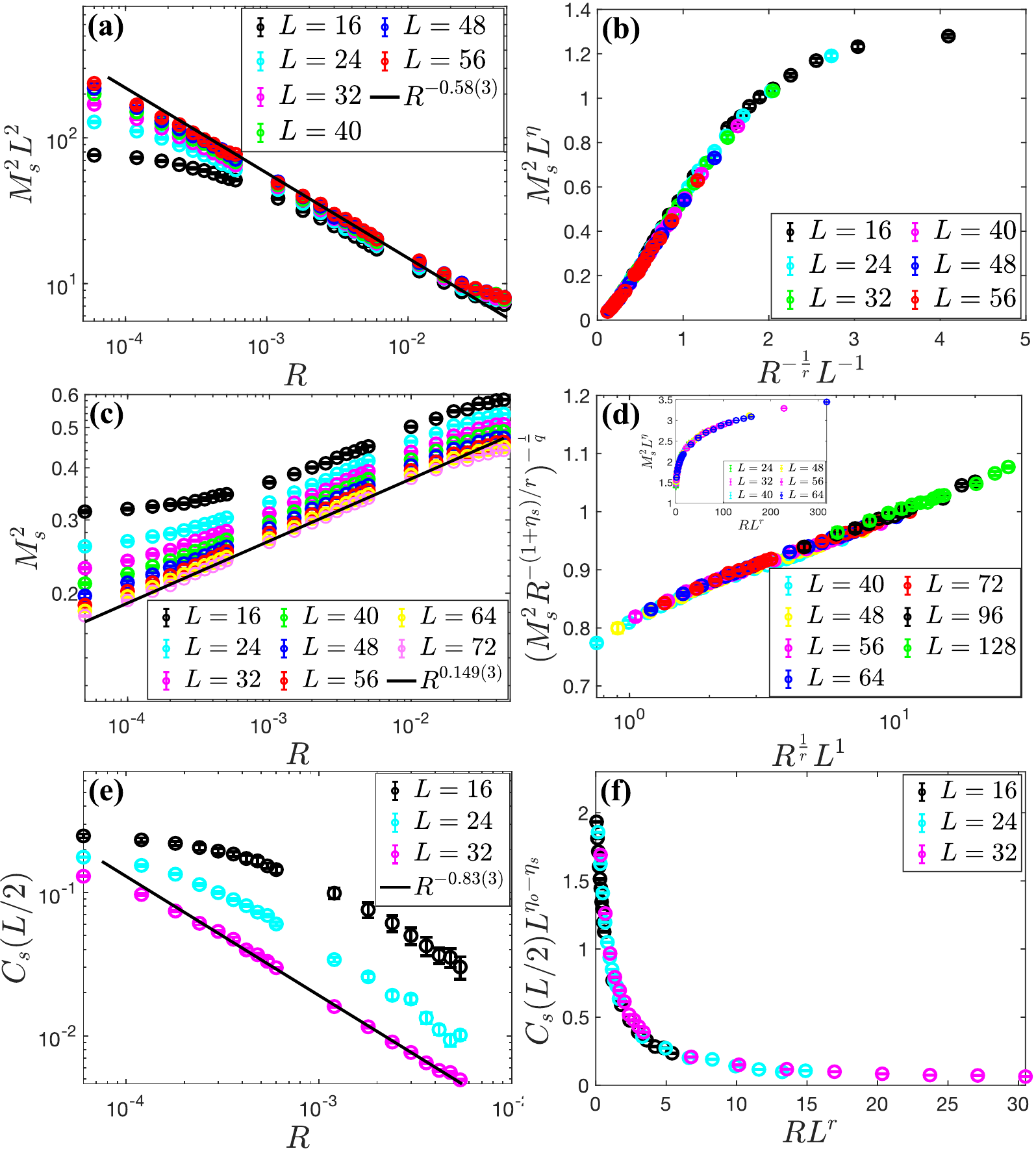}
\caption{This figure presents a finite-time scaling analysis of $M^{2}_{s}$	under non-equilibrium conditions induced by quenching the surface coupling $J_{s}$. Panel (a) corresponds to quenching from the ordinary transition to to the special transition, with the power-law scaling shown as a black solid curve. Panel (b) shown the data collapse for $M^{2}_{s}$ after tunneling $J_{s}$ from ordinary transition to the special point. Panel (c) corresponds to quenching from the extraordinary-log transition to the special transition, where the power-law scaling is also represented by a black solid curve. Panel (d) shows the data collapse for $M^{2}_{s}$ upon tuning $J_{s}$ from extraordinary-log transition to the special transition point. Inset: data collapse at a lower driving rate with Eq.~(\ref{eq:fit_eq}). (e) Decay of the total surface magnetization with system size, showing $M^{2}_{s} \propto L^{-2}$ at a large driving rate of $R=0.045$ for a quench from the ordinary to the special transition point. (f) The spin-spin correlation function $C(L/2)=\left< \vec{S}_{0} \vec{S}_{L/2}\right>$ under the non-equilibrium driven process from the ordinary to the special transition point.} 
\label{fig:ords_exts}
\end{figure}

The driven dynamics of the surface order parameter $M^{2}_{s}$ as a function of the driving rate $R$ are shown in Figs.~\ref{fig:ords_exts}(a) and \ref{fig:ords_exts}(c), respectively. We find that in both cases, the data can be successfully rescaled using Eq.~(\ref{eq:fit_eq}) by simply replacing $r$ with $r_s$, as illustrated in Figs.~\ref{fig:ords_exts}(b) and the inset of (d). This indicates that neither of these critical initial states alters the scaling form of BFTS theory.

In the large-rate limit, for dynamics initiated from the ordinary critical initial state, the order parameter appears to still satisfy the scaling relation derived for the cooling process: the theoretical prediction is $M^{2}_{s} \propto R^{(\eta_s-1)/r_s} \approx R^{-0.616}$, which aligns with our numerical result of $M^{2}_{s} \propto R^{-0.58(3)}$, as shown in the Figs.~\ref{fig:ords_exts}(a). However, this agreement is spurious. For $M^{2}_{s}$, the non-singular (or disordered) contribution, scaling as $M^{2}_{s} \propto L^{-2}$, dominates the initial state fluctuations. Consequently, the singular term, $M^{2}_{s} \propto L^{-(1+\eta_0)} = L^{2.3735}$ (with the ordinary critical anomalous dimension $\eta_0 \approx 1.3735$~\cite{Diehl_1980_Scaling,Diehl_1986_Critical,Wang_2022_Bulk,Zhang_2017_Unconventional}), is masked as a subleading correction. Therefore, to properly analyze the dynamics originating from the ordinary critical initial state, we focus on the two-point surface correlation function at the midpoint, $C_s(L/2)= \left< \vec{S}_{0}  \vec{S}_{L/2} \right>$, where the singular term serves as the dominant contribution: $C_s(L/2) \propto L^{-(1+\eta_0)}$. Consistent with the order parameter analysis, the correlation data can also be successfully rescaled using Eq.~(\ref{eq:fit_eq}) by substituting $M_s^2$ with $C(L/2)$ and replacing $r$ with $r_s$ (see Fig.~\ref{fig:ords_exts}(f)). This further validates the robustness of the FTS form. 

As shown in Fig.~\ref{fig:cool_hat}(f), the data for $C(L/2)$ are well described by a power-law decay, with the black solid line representing a fit yielding $C(L/2) \propto R^{-0.83(3)}$. However, in the large-rate limit, $C(L/2)$ exhibits a significant dependence on the system size. Analogous to Ref.~\cite{Shu_2025_Universal}, introducing a power-law size dependence of the initial state as a correction term can well describes the numerical results: $C(L/2) \propto L^{1+\eta_{o}}R^{(\eta_{s}-\eta_{o})/r_s} \propto R^{-0.772}$.

For the ordered initial state, the FTS gives the scaling relation of $M^2\propto R^{(1+\eta)/r}$~\cite{Zhong2005Dynamic,Huang2014Kibble,Feng2016Theory,Shu_2025_Universal}.
For the driven dynamics by decreasing the surface coupling from the extraordinary-log phase, where the surface correlation function decays as a power of logarithm, a decay even slower than the algebraic decay of long-range order and thus closer to a truly ordered phase, it is quite interesting to study whether this scaling relation for the ordered initial state is still right, we first present the scaling relations derived above for the cooling and heating processes, with the only modification being the replacement of $r$ by $r_s$: $M^{2}_{s} \propto R^{(1+\eta_s)/r_s} \approx R^{0.221}$ and $M^{2}_{s} \propto R^{(\eta_s-1)/r_s} \approx R^{-0.616}$. In the extraordinary-log critical state, $M_s^2$ exhibits an unconventional logarithmic decay, $M^{2}_{s}\propto [\log(L)]^{-q}$ (with $q\approx2.1$), which is distinct from conventional power-law scaling~\cite{Toldin_2021_Boundary}. Fig~\ref{fig:ords_exts}(c) displays $M^{2}_{s}$ as a function of the driving rate $R$. We find that the curve satisfies $M^{2}_{s} \propto R^{0.149(3)}$. We find that the exponent deviates remarkably from $(1+\eta_s)/r_s$. These results apparently demonstrate that the BFTS relation from the ordered phase cannot capture the dynamic scaling from the extraordianry-log phase.

We therefore ascertain that the ultimate scaling relation must incorporate the logrithmic scaling contained in the initial state. In the following, we employ both theoretical analysis and numerical simulations to demonstrate that the extraordinary-log critical initial state induces a novel scaling relation in the large-rate limit.


\textit{\color{blue} Theoretical analysis.---}We begin our analysis from the general FTS framework for $M_s^2$ at $g=0$~\cite{Zhong2005Dynamic,Huang2014Kibble,Feng2016Theory,Zeng_2025_Finite,Shu_2025_Universal}:
\begin{equation}
M_s^2(R, L, m^2_0) = b^{-(1+\eta_s)} M_s^2\left[ R b^{r_s},\, L b^{-1},\, U(m^2_0,b) \right],
\label{mstran}
\end{equation}
in which $U(m^2_0,b)$ is the scale transformation of initial $m^2_0$, and $b$ is the rescaling factor. Then, for large $R$, $R$ dominates and one can obtain the scaling form of $M_s^2$ as
\begin{equation}
M_s^2(R, L, m^2_0) = R^{(1+\eta_s)/r_s} \mathcal{M}_1\left[ L R^{1/r_s},\, U(m_0^2,R^{-1/r_s}) \right],
\label{mstran1}
\end{equation}
by choosing $b=R^{-1/r_s}$ in Eq.~(\ref{mstran}). The dimensionless scaling function $\mathcal{M}_1(x)$, with $x = L R^{1/r_s}$, describes the crossover between two limiting regimes: the equilibrium scaling limit with small $R$ and the KZ scaling limit with large $R$. 

For the dynamics of driving the system across the critical point at a relatively large $R$, the universal information of the initial state can be remembered~\cite{Liu_2014_Dynamic,Huang2014Kibble,Zeng_2025_Finite,Shu_2025_Universal,Wang2026Non-commutative}. Thus, we have $M_s^2\propto[\log(L)]^{-q}$. Since $\mathcal{M}_1(x)$ is a dimensionless scaling function, we deduce that $\mathcal{M}_1(x) \propto [\log(L R^{1/r_s})]^{-q}$. Consequently, for dynamics initiated from the extraordinary-log critical state in the large-rate limit, we obtain the following novel scaling relation:
\begin{equation}
    \begin{split}
    M^{2}_{s} \propto R^{(1+\eta_{s})/r_s}[\log(L R^{1/r_s})]^{-q}
    \end{split}
    \label{eq:exts_FTS}
\end{equation}

We confirm this derivation with our numerical results, as shown in Fig.~\ref{fig:ords_exts}(d). We rescale the axes as follows: the $y$-axis to $[M_s^2 R^{-(1+\eta_s)/r_s}]^{-1/q}$ and the $x$-axis to $\log(R^{1/r_s}L )$. This yields an excellent data collapse over a wide range of system sizes ($L=40$ to $128$). In the figure, we maintain a linear scale for the vertical axis and a logarithmic scale for the horizontal axis; the linear behavior of $\mathcal{M}_1(x)$ in the large-rate regime validates our derivation. 

\textit{\color{blue} Conclusion and discussions.---}In summary, we have systematically investigated the driven critical dynamics near the special surface transition of the three-dimensional Heisenberg model. By implementing distinct nonequilibrium driving protocols, including linearly tuning either temperature or the surface coupling, we have uncovered a rich spectrum of scaling responses. For temperature-driven protocols, both heating and cooling dynamics obey the standard BFTS predictions, consistent with the generalized KZ formalism. The driven dynamics initiated from the ordinary surface critical state also conform to the BFTS framework, with a power-law correction akin to that reported in the Ising mode ~\cite{Shu_2025_Universal}.

Our central finding emerges when the system is driven from the extraordinary-log critical state to the special transition. Here, the logarithmic decay of the initial correlation function, a hallmark of the extraordinary-log universality class, fundamentally alters the large-rate scaling limit. In this case, the BFTS gives a new scaling relation in which the the surface order parameter no longer follows a simple power law but instead acquires a logarithmic term, conforming to the novel scaling relation \(M_s^2 \propto R^{(1+\eta_s)/r_s}[\log(L R^{1/r_s})]^{-q}\). This result underscores that the memory of a non-power-law initial state persists into the driven regime, implying that boundary criticality in extraordinary-log phase yields a richer set of novel dynamical scaling behaviors going beyond the usual KZ mechanism.

\textit{\color{blue} Acknowledgement.---}
S. Y. is supported by the National Natural Science Foundation of China (Grants No. 12222515 and No. 12075324), Research Center for Magnetoelectric Physics of Guangdong Province (Grant No. 2024B0303390001), the Guangdong Provincial Key Laboratory of Magnetoelectric Physics and Devices (Grant No. 2022B1212010008), and the Science and Technology Projects in Guangzhou City (Grant No. 2025A04J5408).
Z.W. thanks the China Postdoctoral Science Foundation under Grants No.2024M752898. The work is supported by the Scientific Research Project (No.WU2025B011), Feng-Ying Career Development Chair Fund and the Start-up Funding of Westlake University.
The authors thank the high-performance computing center and IT office of Westlake University for providing HPC resources and helps.

\bibliographystyle{apsrev4-1}
\bibliography{bibliography}
\clearpage
\appendix
\setcounter{equation}{0}
\setcounter{figure}{0}
\renewcommand{\theequation}{S\arabic{equation}}
\renewcommand{\thefigure}{S\arabic{figure}}
\setcounter{page}{1}

\end{document}